\documentclass[onecolumn]{aa}
\usepackage{graphicx}
\usepackage{graphics}
\usepackage{epsfig}
\usepackage{txfonts}
\usepackage{subfigure}
\usepackage{latexsym}
\begin{document}
%
%
%
\title{On the spherical collapse model in dark energy cosmologies}
\author{David F. Mota\thanks{mota@astro.ox.ac.uk} 
\and Carsten van de Bruck\thanks{cvdb@astro.ox.ac.uk}} 
\institute{Astrophysics, University of Oxford, Keble Road, OX1 3RH, UK}
\abstract{
We study the spherical collapse model in dark energy cosmologies, in which 
dark energy is modelled as a minimally coupled scalar field. We first
follow the standard assumption that dark energy does not cluster on the 
scales of interest. Investigating four different popular potentials in detail, 
we show that the predictions of the spherical collapse model depend on the 
potential used. We also investigate the dependence on the initial conditions. 
Secondly, we investigate in how far perturbations in the quintessence field 
affect the predictions of the spherical collapse model. In doing so, we 
assume that the field collapses along with the dark matter. Although 
the field is still subdominant at the time of virialisation, the predictions 
are different from the case of a homogeneous dark energy component. 
This will in particular be true if the field is non--minimally coupled.
We conclude that a better understanding of the evolution of dark energy 
in the highly non--linear regime is needed in order to make predictions 
using the spherical collapse model in models with dark energy.
}
\date{Received: xxx; accepted: xxx}
\maketitle
\keywords{Cosmology: theory -- Cosmology: large-scale structure of Universe -- Cosmology: miscellaneous}
                
\section{Introduction}

The most surprising result of observational cosmology is the discovery 
that high redshift supernovae are less bright than expected \cite{perlmutter,tonry}. 
The interpretation of this finding is that the expansion of the universe 
accelerates, instead of slowing down. The matter responsible for this 
accelerated expansion is dubbed ``dark energy'' and, according to 
General Relativity, needs to have negative pressure. 

One important goal 
of cosmology is to obtain information about the nature of dark energy 
(see e.g. \cite{ratra} for a review). The physical origin of dark energy 
is unknown and its likely that new physics beyond the standard model 
of particle physics will be needed to explain its properties. 
Scalar fields are plausible candidates for dark energy (see 
e.g. \cite{wetterich1, wetterich11, ratra1,  wetterich12}) (these scalar 
fields are called quintessence fields). A distinctive 
feature of these models is that the properties of dark energy, such as its equation 
of state, generally vary during the cosmic history. 

Cosmological observations probe the properties of dark energy.
The anisotropy spectrum of the Cosmic Microwave Background (CMB), the shape 
of matter power spectrum and the distance-redshift relation are possible 
sources of information and will  
help us to distinguish between the different models of dark energy. 

The behaviour of small perturbations in a scalar field and its effect on CMB 
anisotropies and structure formation has been investigated by a number of 
authors. However, the behaviour of quintessence during the gravitational 
collapse into the highly non-linear regime is not well understood and currently 
under investigation (see e.g. \cite{bartelmann, wetterich21, dolag,
  maccio1, maccio2, mainini1,   mainini2, matarrese, perrotta,
  amendola2, linder, guzman, guz} for recent work).  

Usually it is assumed that there are no density fluctuations in the quintessence 
field on cluster scales and below. The reason for this assumption is that, according to 
linear perturbation theory, the mass of the field is very small (the associated 
wavelength of the particle is of the order of the Hubble radius) and, hence, it 
does not feel overdensities of the size of tenth of Mpc or smaller \cite{wang}.
Let us assume for instance the particle physics
candidate for dark energy, a scalar field $\phi$. 
In the linear regime of the small cosmological
perturbations, it
can be shown that, during the matter dominated epoch, perturbations in
the scalar field are described by \cite{hwang}
\begin{equation}
\delta\ddot\phi+3H\delta\dot\phi+(k^2/a^2+V'')\delta\phi=\dot\phi\delta_{cdm}
\label{linear}
\end{equation}
where the metric was perturbed about a flat Friedmann-Robertson-Walker (FRW)
metric.  From equation (\ref{linear}) the effective Jeans length for linear
perturbation of the scalar field is roughly given by 
\begin{equation}
\lambda_{\rm J} \approx 2\pi/\sqrt{V_{,\phi\phi}} 
\end{equation}
which turns out to be of the order of the horizon size \cite{ma}. Thus, on scales
much smaller than $\lambda_{\rm J}$ the  fluctuations on the field 
are unimportant. This results lead people to assume that dark energy
is an homogeneous field throughout the universe, which is not affected
by structure formation on small scales.  However, the
source term $\dot\phi\delta_{cdm}$ implies that
dark energy is not smoothly spread 
\cite{caldwell}. And in spite of the perturbations in 
dark energy are present and grow  only on scales about the horizon size and
larger, the effects on the evolution of matter overdensities  are indeed important
and significant \cite{ferreira}. For instance, the spatially inhomogeneous
distinctive property of quintessence, with respect to the cosmological
constant $\Lambda$, has an important effect on the large-angular-scale
of the CMBR \cite{steinhardt1}. To study the effect of dark energy on small scales, 
N--body simulations were used and the assumptions made that quintessence perturbations 
are negligible. (In \cite{maccio2}, a possible coupling of dark energy and dark 
matter was explicitly taken into account for the perturbations.) 

The assumption of neglecting the effects of matter perturbations on the
evolution of dark energy (and its backreaction) at small scales is indeed a good
approximation when perturbations in the metric are 
very small. Notice however, if the field couples explicitly to matter, such 
as in the {\it coupled quintessence} scenario (see e.g. \cite{wetterich12} 
and \cite{amendola1}), the conclusion that the Jeans length is of order the 
horizon size should no longer hold.

Nevertheless, even if the field is not coupled, one should be careful when extrapolating 
the small-scale linear-regime results to the highly non-linear regime. Then, locally the flat FRW
metric is not a good approximation anymore to describe the geometry of
overdense regions. It is natural to think that once 
a dark matter overdensity decouples from the background expansion and collapses, the field 
inside the cluster feels the gravitational potential inside the overdensity and 
its evolution will be different from the background evolution. 
The backreaction effects in the highly non-linear regime could influence 
the evolution of perturbations in dark energy considerably, which in turn 
influence the evolution of the matter perturbation. In \cite{magueijo} it 
was suggested that the quintessence field could have an important impact 
in the highly non--linear regime. 
Due to these considerations, one may even ask if the quintessence field 
can be important even on galactic scales? It was speculated by 
\cite{wetterich2,wetterich21} and by \cite{Arbey} that 
highly non--linear perturbations might indeed be important even on 
galactic scales. It was found that, at least in principle, the quintessence (or 
a scalar field) field could be responsible for the observed flat rotation curves in
galaxies. In \cite{pad1,pad2,bag} and \cite{causse} more exotic models, based on tachyon fields, 
have been discussed and it was argued that the equation of state is scale--dependent.

If it turns out that backreaction effects of metric and density perturbations 
in dark matter could influence perturbations of quintessence on small scales, 
this could significantly change our understanding of structure formation on
galactic and cluster scales in models with quintessence. 

One popular model to study the non--linear growth in cold dark matter is the 
spherical collapse model (see. e.g. \cite{Padmanabhan} or \cite{peacock}). 
The model was used first in the standard cold dark 
matter scenario, but later also in the 
cold dark matter model with cosmological constant ($\Lambda$CDM) by
\cite{lahav}. Recently, 
the model was also used in quintessential scenarios and dark matter under the 
assumption that dark energy does not cluster on scales much smaller than the 
horizon (see e.g. \cite{wang}). It was then subsequently used in order to 
make predictions for cosmological observations 
(see e.g. \cite{Weller}, \cite{weinberg}, \cite{battye}).

The aim of this paper is two--fold. Firstly, we model dark energy as a non--minimally 
coupled scalar field and consider four different 
potentials in order to investigate how the predictions of the spherical 
collapse model depend on the potential. In addition, the effect of the initial conditions 
are investigated as well. Our second aim is to get a feel about the effects of perturbations 
in the quintessence field on the predictions of the spherical collapse model. 
In doing so, we investigate two extreme 
cases: in the first case we follow the literature and assume that there are no 
fluctuations in the quintessence field on the scales of interest. We then assume the 
other extreme case and assume that the field inside the overdensity collapses 
together with the dark matter\footnote{This approach was first used, in 
the context of varying-fine structure theories \cite{mota1, mota2}.}.
As we will see, the predictions of the 
spherical collapse model will be quite different in this case, although 
the field is only slightly non--linear (with density contrasts of order one) 
at the time of virialisation, whereas dark matter is in the highly non--linear regime.  
Our results imply that the predictions of the spherical collapse model 
depend on the assumptions made for the clustering properties of the 
quintessence in that model. Thus, a better understanding of the behaviour of the scalar field 
in the highly non--linear regime is needed, if the spherical collapse model is 
used for predictions involving large scale structures, such as cluster abundances, 
weak and strong lensing, etc.

The paper is organised as follows: In Section 2 we describe briefly the 
spherical collapse model and write down the equations used. We also 
give an overview of the potentials used. In Section 3 we describe our 
assumptions and give the results of our numerical calculations. 
A discussion of the results and our conclusions can be found in Section 4.

\section{The spherical collapse model}
We consider a flat, homogeneous and isotropic background universe with 
scale factor $a(t)$. Since we are interested in the matter dominated epoch, 
when structure formation starts, we assume that the universe is filled with cold
dark matter of density $\rho _{m}$ $\propto a^{-3}$ and a 
dark energy fluid, with energy-density $\rho _{\phi}$. 
The equations that describe our background universe
are (we set $\hbar =c\equiv 1$ throughout the paper):
\begin{eqnarray}
3H^{2}&=&8\pi G\left( \rho _{m}+\rho _{\phi }\right)  
\label{fried}\\
\dot\rho_{\phi}&=&-3 H (1+w_{\phi})\rho_{\phi}
\label{psidot}
\end{eqnarray}
where $H\equiv \dot{a}/a$ is the Hubble rate. 
When $w_{\phi}=-1$ then dark energy is the
vacuum energy density $\Lambda$.
If dark energy is a scalar field $\phi$ (quintessence), 
$\rho_{\phi}=\frac{1}{2}\dot\phi^2+V(\phi)$ and
$P_{\phi}=\frac{1}{2}\dot\phi^2-V(\phi)$, where $V(\phi)$ is the
scalar field potential. In this case, it is useful to re-write
equation (\ref{psidot}) as
\begin{equation}
\ddot\phi+3 H \dot\phi+V^{'}=0
\label{phidot}
\end{equation}
where the prime represents a derivative with respect to $\phi$.
In this paper we consider four examples of quintessential potentials:
\begin{itemize}
\item The double exponential potential \cite{copel}:
\begin{equation}
V(\phi) = M\left(\exp(\beta\phi)+\exp(\gamma\phi)\right).
\end{equation}
\item The exponential potential with inverse power \cite{steinhardt}:
\begin{equation}
V(\phi) = M\left(\exp(\gamma/\phi)-1\right)
\end{equation}
\item The Albrecht-Skordis model \cite{skordis}:
\begin{equation}
V(\phi) = M\left(A + \left(\phi - B\right)^2\right)\exp(-\gamma\phi)
\end{equation}
\item The supergravity--motivated potential \cite{brax}:
\begin{equation}
V(\phi) = M\exp(\phi^2)/\phi^\gamma
\end{equation}
\end{itemize}
We choose the parameters in the potentials and the initial conditions in 
the background such that today we obtain $\Omega_m=0.3$, $\Omega_{\phi}=0.7$,
$H_0=100\,h\,$km/(Mpc$\cdot$ sec) with $h=0.7$ and $-1\leq w_{\phi}\leq-0.8$.

In order to study the non--linear evolution of the gravitational 
collapse we make use of the spherical collapse (or infall) model. Here, one considers 
a spherical overdense region of radius $R$ and models the interior spacetime as a 
FRW universe. 
This approach is equivalent to study the effect of perturbations to the
Friedmann metric by considering spherically symmetric regions of different
spatial curvature in accord with Birkhoff's theorem. Clearly, this
model ignores any anisotropic effects of gravitational instability or
collapse. However, the model is quite useful to learn about non--linear 
gravitational collapse and it is widely used as a starting point for
(semi-) analytical models of large-scale structure formation. 

Consider a spherical perturbation, in the cold dark matter fluid, 
with (spatially) constant internal density which, at
an initial time, has an overdense amplitude $\delta _{i}>0$ and $|\delta
_{i}|\ll 1$. The cold dark matter density inside the cluster is then initially
$\rho_{cdm}=\rho_{m}(1+\delta(R_i,t_i))$, where $R_i$ is the initial
radius of the overdensity. The dark energy fluid density inside the overdensity,
$\rho_{\phi_c}$,  will initially be considered the same as the
background one\footnote{This assumption will be made throughout the paper.}.
At early times the sphere expands along with the background, but with a slightly 
different expansion rate. For a
sufficiently large $\delta _{i},$ gravity prevents the sphere from
expanding for ever: at one point, the overdensity will stop expanding and 
start to collapse. Three characteristic phases during the evolution 
can then be identified:
\begin{itemize}
\item \textit{Turnaround}: the sphere breaks away from the general expansion 
and reaches a maximum radius. 
\item \textit{Collapse}: if only gravity is significant, the
sphere will then collapse towards a central singularity where the densities
of the matter fields would formally go to infinity. In practise, pressure
and dissipative physics intervene well before this singularity is reached
and convert the kinetic energy of collapse into random motions. 
\item \textit{
Virialisation}: dynamical equilibrium is reached and the system becomes
stationary: the radius of the system the energy of the different components 
are constant. 
\end{itemize}

In  the spherical collapse model, due to its symmetry, 
the only independent coordinates are
the radius of the overdensity and time.  Also, as is standard practise
when using this model, we consider that there are no shell-crossing 
which implies mass (of cold dark matter) conservation inside the overdensity 
and independence of the radius coordinate \cite{Padmanabhan}. The equations 
can then be written ignoring the spatial dependence of the fields (but still 
including an equation for the evolution of the radius). 
The evolution of a spherical overdense
patch of scale radius $R(t)$ is given by the Raychaudhuri equation: 
\begin{equation}
3\ddot{R}=-4\pi G R\left( \rho _{\rm cdm}+\rho _{\phi_c}(1+3w_{\phi_c})\right)  
\label{rcluster}
\end{equation}
Note that it would be a wrong to use the Friedmann equation
for a closed universe with a constant curvature $k$, since the former can 
vary in time \cite{weinberg, wang}.
In the cluster, the evolution of $\rho_{\phi _{c}}$ and $\rho _{\rm cdm}$ 
is given by 
\begin{equation}
\dot\rho_{\phi_c}=-3\frac{\dot R}{R}(1+w_{\phi_c})\rho_{\phi_c} + \Gamma
\label{psidotcluster}
\end{equation}
and $\rho _{cdm}\propto R^{-3}$ due to mass conservation. The quantity 
$\Gamma$ describes the energy loss of dark energy inside the dark matter 
overdensity, as this component does not necessary follow the collapse of 
dark matter and energy can formally flow out of the system. As such, $\Gamma$ encodes 
in how far backreaction effects from the dark matter non--linearities act on 
the dark energy component. Formally, $\Gamma$ can also describe the coupling between 
matter and dark energy, in which case we would have to add a term containing 
$\Gamma$ in the energy conservation equation for dark matter. In this case it 
would be no longer true that $\rho_{\rm cdm} \propto R^{-3}$.

Once again, in the case of a scalar field, equation (\ref{psidotcluster}) 
can be written as
\begin{equation}
\ddot{\phi _{c}}+3\frac{\dot{R}}{R}\dot{\phi _{c}}+V_c{'}(\phi_c)=
\frac{\Gamma}{\dot\phi} 
\label{phidotcluster}
\end{equation}
where $\phi_c$ is the field inside the overdensity in order to distinguish 
it from the background value and $V_c=V(\phi_c)$ is its potential.

We will evolve the spherical overdensity from high redshift until its
virialisation occurs. According to the virial theorem, equilibrium will be
reached when 
$T=\frac{1}{2}R\frac{\partial U}{\partial R}$; 
$T$ $=\frac{1}{2}M\bar{v}_{vir}^{2}$ is the total kinetic energy
at virialisation and $\bar{v}_{vir}^{2}$ is the mean-square velocity of the
components of the cluster, and $U$ is the average total potential energy
in the sphere. It is useful to write the condition for virialisation to occur in
terms of the potential energies associated the different components of the
overdensity. The potential energy for a given component $^{\prime }x^{\prime
}$ can be calculated from its general form in a spherical region \cite{
landaubook}: 
\begin{eqnarray}
U_{x} &=&2\pi \int_{0}^{R}\rho _{tot}\Phi _{x}r^{2}dr, \\
\Phi _{x}\left( r\right) &=&-2\pi G\rho _{x}(1+3w_x)\left( R^{2}-\frac{r^{2}}{3}%
\right) ,  \label{generalpotential}
\end{eqnarray}%
where $\rho _{tot}$ is the {\it total} energy density inside the sphere, 
$\Phi_{x} $ is the gravitational potential due to component $x$ with 
energy density $\rho_{x}$ and equation of state $w_x$. In our
calculation, we have taken also the pressure 
contribution from dark energy into account, since the Poisson--equation
reads (including the correction from General Relativity)
\begin{equation}
\Delta \Phi = 4\pi G\left(\rho + 3 P \right).
\label{poisson}
\end{equation}

Using the virial theorem and energy conservation at the
turnaround and cluster virialisation times, we obtain an equilibrium
condition only in terms of the potential energies: 
\begin{equation}
\frac{1}{2} R \frac{\partial}{\partial R} (U_{G}+ U_{\phi_c})+
U_{G }+U_{\phi_{c}}\,\vert_{z_{v}}=U_{G}+U_{\phi _{c}}\,\vert_{z_{ta}}
\label{virialcond}
\end{equation}
where $U_{G}$ is the potential energy for Cold Dark Matter,
$U_{\phi_c}$  is the potential energy for dark energy, 
$z_{v}$ is the redshift of virialisation and $z_{ta}$ is the redshift
at the turnaround of the over-density at its maximum radius, when $R=R_{max}$
and $\dot{R}=0$. In the case where $w_{\phi}=-1$ the expressions reduce to the 
usual virialisation condition for $\Lambda$CDM models. One should
point out here the inconsistency widely used in the literature, when
one makes use of equation (\ref{virialcond})
together with the assumption that energy is not conserved inside the
overdensity (dark energy is assumed to be homogeneous, so it has to
flow out from the overdensity). In the case where
$\Gamma=0$ the whole process of virialisation is self-consistent.

In addition, we also discuss the linear growth factor 
$D(z) = (\delta_{\rm c}(z)/a(z))/(\delta_{\rm c}(0)/a(0)) $ 
for the four different potentials. The linear density contrast $\delta_{\rm c}$ obeys 
the equation 
\begin{equation}
\ddot \delta_{\rm c} + 2H \dot \delta_{\rm c} - 4\pi G \rho_{\rm m} \delta_{\rm c} = 0.
\end{equation}
As we will see, the growth factor $D(z)$ depends on the potential used.

The behaviour of dark energy during the evolution of a cluster can now be 
obtained by numerically evolving the background eqs.(\ref{fried})-(\ref{phidot}) 
and the cluster eqs.(\ref{rcluster})-(\ref{phidotcluster}) until the 
virialisation condition (\ref{virialcond}) holds.  Additionally to 
the four potentials mentioned above, we also have considered three different 
types of large scale structure models: the Standard Cold Dark Matter model 
$SCDM$, the $\Lambda$CDM model, and a model with dark energy with constant 
equation of state $w_{\phi}=-0.8$ ($QCDM$). The SCDM and the $\Lambda$CDM 
model was used to test our code and we will not give the results for these 
models, as they are well known. 

In next section we will make two assumptions about the function $\Gamma$.
In one case we assume that the field is homogeneous all over the Universe, 
so that the field inside the overdensity evolves just like the background. In the second case we 
assume that the field collapses together with the dark matter. We would like 
to point out that our assumptions are different from the ones 
made in the first version of \cite{Lokas}, where it was assumed that dark energy 
is homogeneous but a Friedmann--like equation was used to study the evolution of $R(t)$.

\section{Non--linear collapse with dark energy}
We now study numerically the equations given in the last section. 
For this, we must make assumptions about the function $\Gamma$, which 
encodes the backreaction of dark matter perturbations on dark energy.
We will make two assumptions for $\Gamma$: in the first case, 
the quintessence field is assumed to be smooth throughout space. 
This can be obtained by putting 
\begin{equation}\label{gammamodel}
\Gamma = -3\left(\frac{\dot a}{a}-\frac{\dot R}{R}\right)\dot\phi_c^2 
\end{equation}
and $\phi_c(t_i) = \phi(t_i)$ and $\dot\phi_c(t_i)=\dot\phi(t_i)$.
Then we obtain $\phi_c(t) = \phi(t)$ for all times. This is the standard
assumption made in the literature. As already mentioned in the 
introduction, the main reason for this assumption is that
the effective Jeans length for linear perturbations turns out 
to be of the order of the horizon size. Thus, so the 
argument goes, on scales much smaller than $\lambda_{\rm J}$ the 
fluctuations are unimportant.

The form of $\Gamma$ chosen above is clearly not the 
full truth. At least at very late times during the collapse of the 
dark matter, especially when the density contrast in dark matter is very large 
($\delta_{\rm m}\gg1$), 
the field should no longer feel the background metric,
i.e. expand with the background, but decouple from it. In 
this regime, the evolution of dark energy could be different and influence the 
details of the collapse.
The detail of this can only be obtained from a fully relativistic calculation, 
which is beyond the scope of this paper. However, to get a feel of what
may happen we make now the other extreme assumption that the field follows 
the dark matter collapse from the very beginning. That is, we assume that 
$\Gamma = 0$. This is clearly not a realistic assumption either, but the results 
we obtain are surprising and interesting in itself. 

\subsection{Results for the four quintessence potentials}
\begin{figure*}[Hhtbp!]
\centering
\includegraphics[height=10cm]{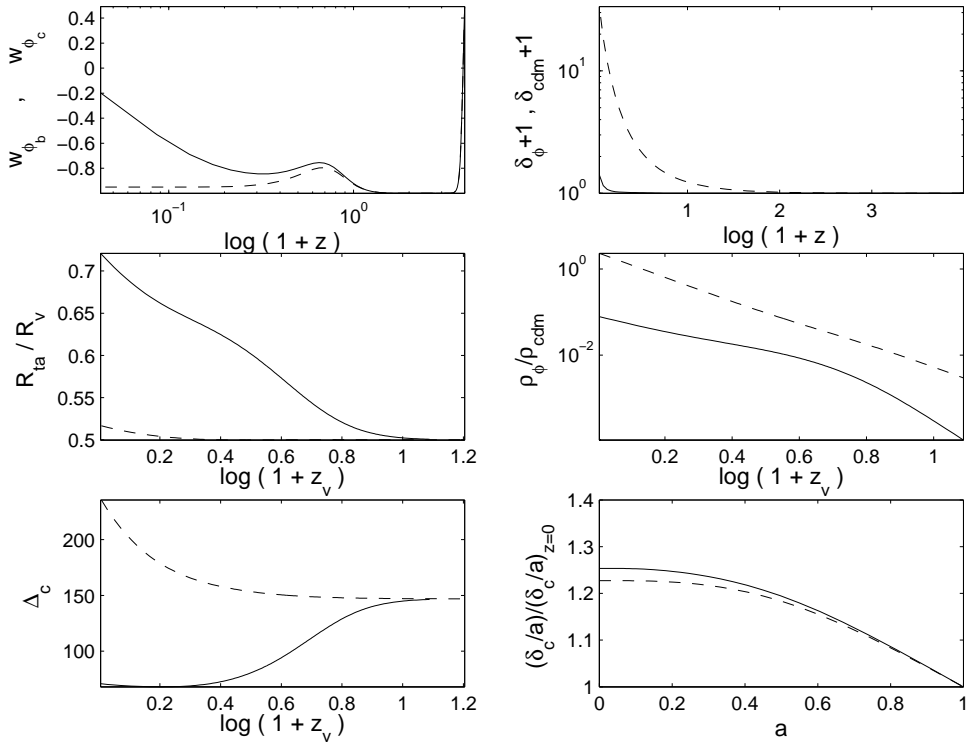}
\caption{{\protect\small{\textit{Quintessence model 
$V=M(\exp(\beta\phi) + \exp(\gamma\phi))$ \cite{copel}. Top left panel:
Evolution of $w_{\phi}$ in the background (dashed line) and inside
an overdensity (solid line) for $\Gamma=0$, as a function of $\log (1+z)$
(overdensity virialises at $z=0$). 
Top right panel: Evolution of $\rho_{\phi}/\rho_{\phi_c}$  (solid
line) and $\rho_{m}/\rho_{cdm}$ (dashed line) in the case $\Gamma=0$ 
as a function of $\log(1+z)$ (overdensity virialises at $z=0$). Middle left panel: The ratio 
$R_{ta}/R_V$ in the case of an inhomogeneous ($\Gamma=0$) (solid line) 
and homogeneous scalar field (dashed line). 
The ratio $\rho_\phi/\rho_{\rm matter}$ inside the overdensity 
(solid line) ($\Gamma=0$) and in the background (dashed line). 
Bottom left: $\Delta_{c}$ as a function of $z_V$, considering the effect
of an inhomogeneous quintessence field ($\Gamma=0$) (solid line). The results 
for a homogeneous quintessence field are shown as well (dashed line). 
Bottom right:  Predictions for the linear growth factor for this potential (solid line).
The dashed line represents the $\Lambda$CDM case.}}}}
\label{copel}
\end{figure*}

\begin{figure*}[Hhtbp!]
\centering
\includegraphics[height=10cm]{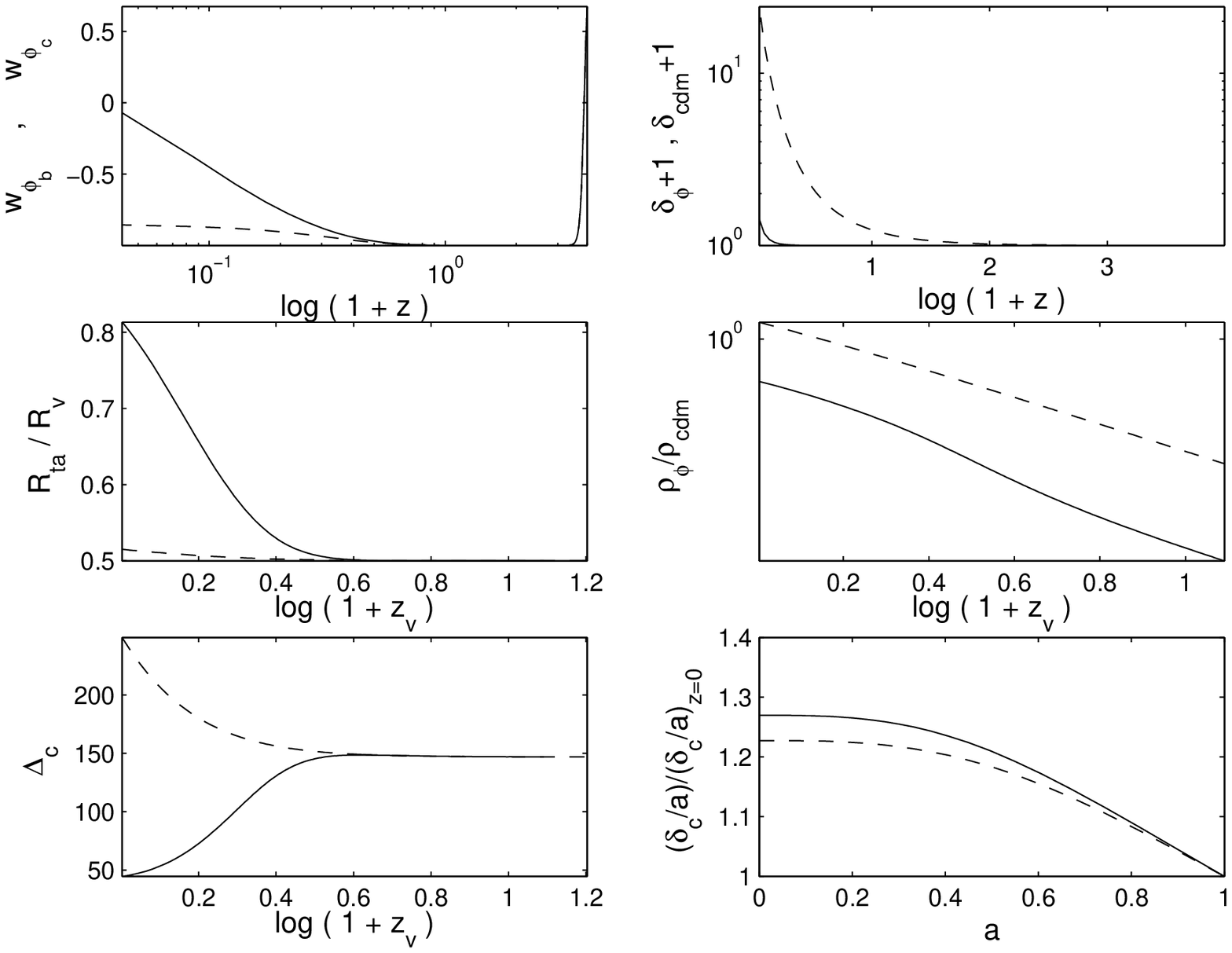}
\caption{{\protect\small {\textit{Quintessence model $V= M(\exp(\gamma/\phi)-1)$ \cite{steinhardt}. 
Top left panel:
Evolution of $w_{\phi}$ in the background (dashed line) and inside
the overdensity (solid line) for $\Gamma=0$ as a function of 
$\log (1+z)$ (overdensity virialises at $z=0$). 
Top right panel: Evolution of $\rho_{\phi}/\rho_{\phi_c}$  (solid
line) and $\rho_{m}/\rho_{cdm}$ (dashed line) in the case $\Gamma=0$ 
as a function of $\log(1+z)$ (overdensity virialises at $z=0$). Middle left panel: The ratio 
$R_{ta}/R_V$ in the case of an inhomogeneous ($\Gamma=0$) (solid line) 
and homogeneous scalar field (dashed line). 
The ratio $\rho_\phi/\rho_{\rm matter}$ inside the overdensity  
(solid line) ($\Gamma=0$) and in the background (dashed line). 
Bottom left: $\Delta_{c}$ as a function of $z_V$, considering the effect
of an inhomogeneous quintessence field ($\Gamma=0$) (solid line). The results 
for a homogeneous quintessence field are shown as well (dashed line). 
Bottom right:  Predictions for the linear growth factor for this potential (solid line).
The dashed line represents the $\Lambda CDM$ case.}}}}
\label{stein}
\end{figure*}

\begin{figure*}[Hhtbp!]
\centering
\includegraphics[height=10cm]{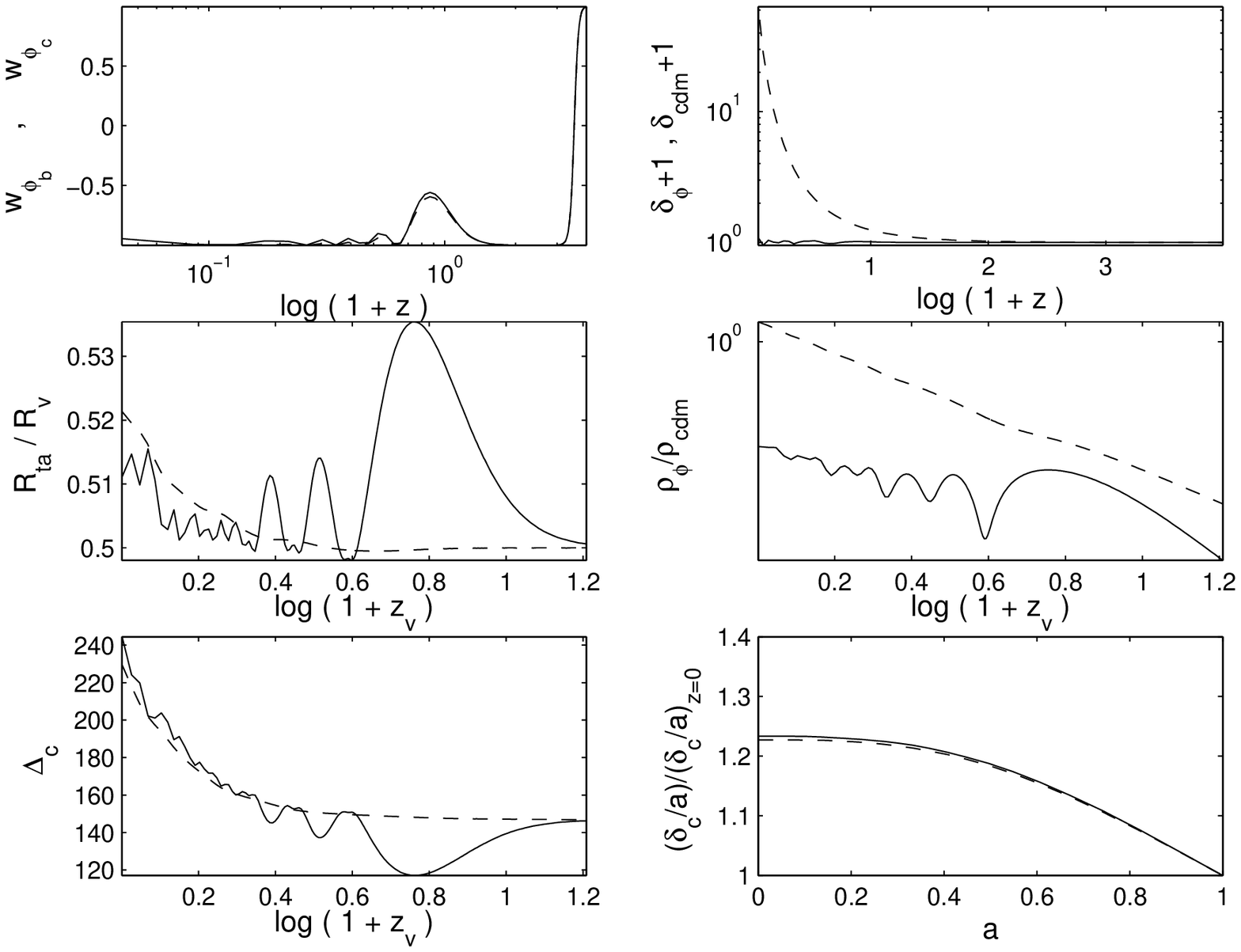}
\caption{{\protect\small {\textit{Quintessence model 
$V= M(A+(\phi-B)^2)\exp(-\gamma\phi)$ \cite{skordis}. 
Top left panel:
Evolution of $w_{\phi}$ in the background (dashed line) and inside the 
overdensity (solid line) for $\Gamma=0$ as a function of $\log (1+z)$ 
(overdensity virialises at $z=0$). 
Top right panel: Evolution of $\rho_{\phi}/\rho_{\phi_c}$  (solid
line) and $\rho_{m}/\rho_{cdm}$ (dashed line) in the case $\Gamma=0$ 
as a function of $\log(1+z)$ (overdensity virialises at $z=0$). Middle left panel: The ratio 
$R_{ta}/R_V$ in the case of an inhomogeneous ($\Gamma=0$) (solid line) 
and homogeneous scalar field (dashed line). 
The ratio $\rho_\phi/\rho_{\rm matter}$ inside the overdensity 
(solid line) ($\Gamma=0$) and in the background (dashed line). 
Bottom left: $\Delta_{c}$ as a function of $z_V$, considering the effect
of an inhomogeneous quintessence field ($\Gamma=0$) (solid line). The results 
for a homogeneous quintessence field are shown as well (dashed line). 
Bottom right:  Predictions for the linear growth factor for this potential (solid line).
The dashed line represents the $\Lambda CDM$ case.}}}}
\label{skordis}
\end{figure*}

\begin{figure*}[Hhtbp!]
\centering
\includegraphics[height=10cm]{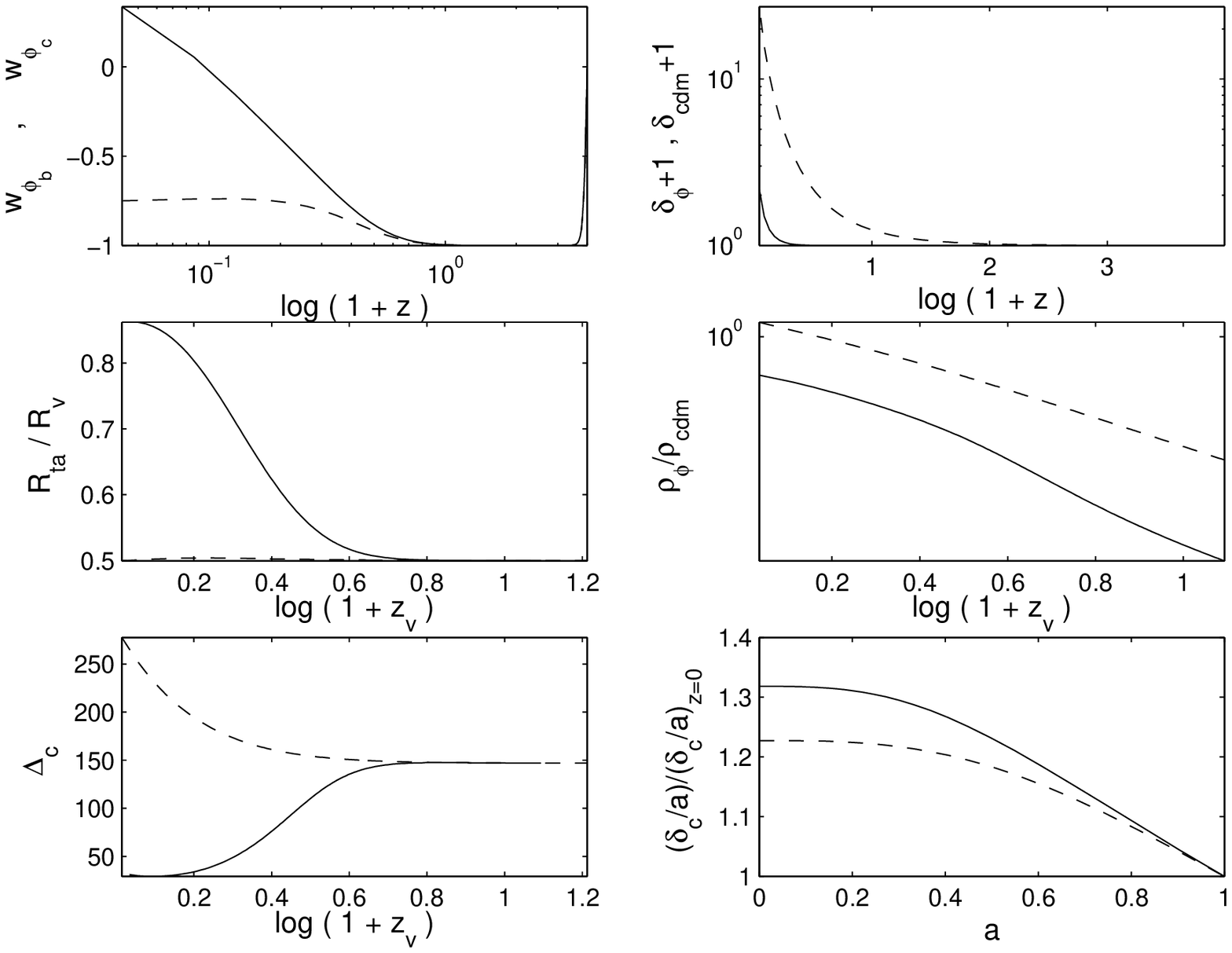}
\caption{{\protect\small {\textit{Quintessence model 
$V= M\exp(\phi^2)/\phi^{\gamma}$ \cite{brax}. 
Top left panel:
Evolution of $w_{\phi}$ in the background (dashed line) and inside the 
overdensity (solid line) for $\Gamma=0$ as a function of 
$\log (1+z)$ (overdensity virialises at $z=0$). 
Top right panel: Evolution of $\rho_{\phi}/\rho_{\phi_c}$  (solid
line) and $\rho_{m}/\rho_{cdm}$ (dashed line) in the case $\Gamma=0$ 
as a function of $\log(1+z)$ (overdensity virialises at $z=0$). Middle left panel: The ratio 
$R_{ta}/R_V$ in the case of an inhomogeneous ($\Gamma=0$) (solid line) 
and homogeneous scalar field (dashed line). 
The ratio $\rho_\phi/\rho_{\rm matter}$ inside the overdensity 
(solid line) ($\Gamma=0$) and in the background (dashed line). 
Bottom left: $\Delta_{c}$ as a function of $z_V$, considering the effect
of an inhomogeneous quintessence field ($\Gamma=0$) (solid line). The results 
for a homogeneous quintessence field are shown as well (dashed line). 
Bottom right:  Predictions for the linear growth factor for this potential (solid line).
The dashed line represents the $\Lambda CDM$ case.}}}}
\label{braxs}
\end{figure*}

We begin our discussion with an analysis of the four different 
quintessential potentials mentioned in Section 2. The results 
of our numerical calculations are presented in 
Fig. \ref{copel}, \ref{stein}, \ref{skordis} and \ref{braxs}.


In these plots we show for all models the behaviour of the equation of 
state inside the cluster and in the background (for the case $\Gamma=0$), 
the density contrast in matter and field (again for $\Gamma=0$), 
the ratio of the radius at turnaround $R_{ta}$ and radius at virialisation 
$R_V$ for both the case $\Gamma=0$ and an homogeneous scalar field, 
the ratio of the energy densities in field and matter, the non--linear 
density contrast for both cases and the growth factor for the models 
compared to the $\Lambda$CDM model. 

As it is clear from the upper right 
plots, even in the $\Gamma=0$ case (ie. the quintessence field collapses 
with together with matter), the field will be subdominant in the virialised 
object. This can also be seen from the Figures on the right hand side in 
the middle panels in Figures  \ref{copel}, \ref{stein}, \ref{skordis}
and \ref{braxs}
: in all cases the ratio 
$\rho_\phi/\rho_{\rm cdm}$ is much smaller than 
one. Well before turnaround, i.e. while the density contrast in dark matter is smaller 
than one, our results agree with what one would obtain from linear perturbation theory. 
For $\Gamma=0$, the perturbations in dark energy remain very small up to this point. 
Even after turnaround, our results imply that fluctuations in the dark energy component 
remain small, while dark matter enters the highly non--linear regime. 

Concerning the clustering properties of the quintessence field, 
it can significantly alter the predictions for the non--linear 
density contrast $\Delta_{\rm c} 
= \rho_{\rm cdm,cluster}(t_{\rm V})/\rho_{\rm cdm,backgr}(t_{\rm V})$ 
at virialisation: assuming that the field 
is inhomogeneous $\Gamma=0$, for small virialisation redshifts the predictions 
can differ by a factor of four or more. In the same way the predictions 
for the ratio $R_{\rm ta}/R_{\rm V}$ strongly depends on how the scalar 
field behaves during the highly non--linear regime. The reason for this 
is, that the just before the overdensity formally becomes a singularity, 
the quintessence field becomes more and more important and changes the 
evolution of $R(t)$ during the last stages of the collapse. We remind the 
reader that, in the Einstein--de Sitter model $\Delta_{\rm c}\approx 147$\footnote{If we would
define $\Delta_{\rm c2} 
= \rho_{\rm cdm,cluster}(t_{\rm V})/\rho_{\rm cdm,backgr}(t_{\rm c})$, where $t_c$ is the 
collapse time, its value in the Einstein--de Sitter model would be approximately 178. However, 
for computational convenience, we consider instead the quantity $\Delta_{\rm c}$ at the time 
of virialisation. $\Delta_{\rm c}$ and $\Delta_{\rm c2}$ differ only by a factor from 
each other.}  
At high virialisation redshifts, all models predict 
$\Delta_{\rm c} \approx 147$, only at low virialisation redshifts, 
significant deviations can be expected due to the fact that the 
dark energy becomes more and more important. Note that in the Albrecht--Skordis 
model, the field behaves like a cosmological constant in the background (see Fig \ref{skordis}). 
Hence, one would expect that the differences to the $\Lambda$CDM model are small. This is 
indeed the case, because $\dot \phi_{\rm c} \approx 0$ and, hence, $\Gamma \approx 0$, as 
it can be seen from eq. (\ref{gammamodel}). Thus, in this model, the 
fluctuations in the quintessence field remain small and the field is 
almost homogeneous. 

\begin{figure*}[Hhtbp!]
\centering
\includegraphics[height=10cm]{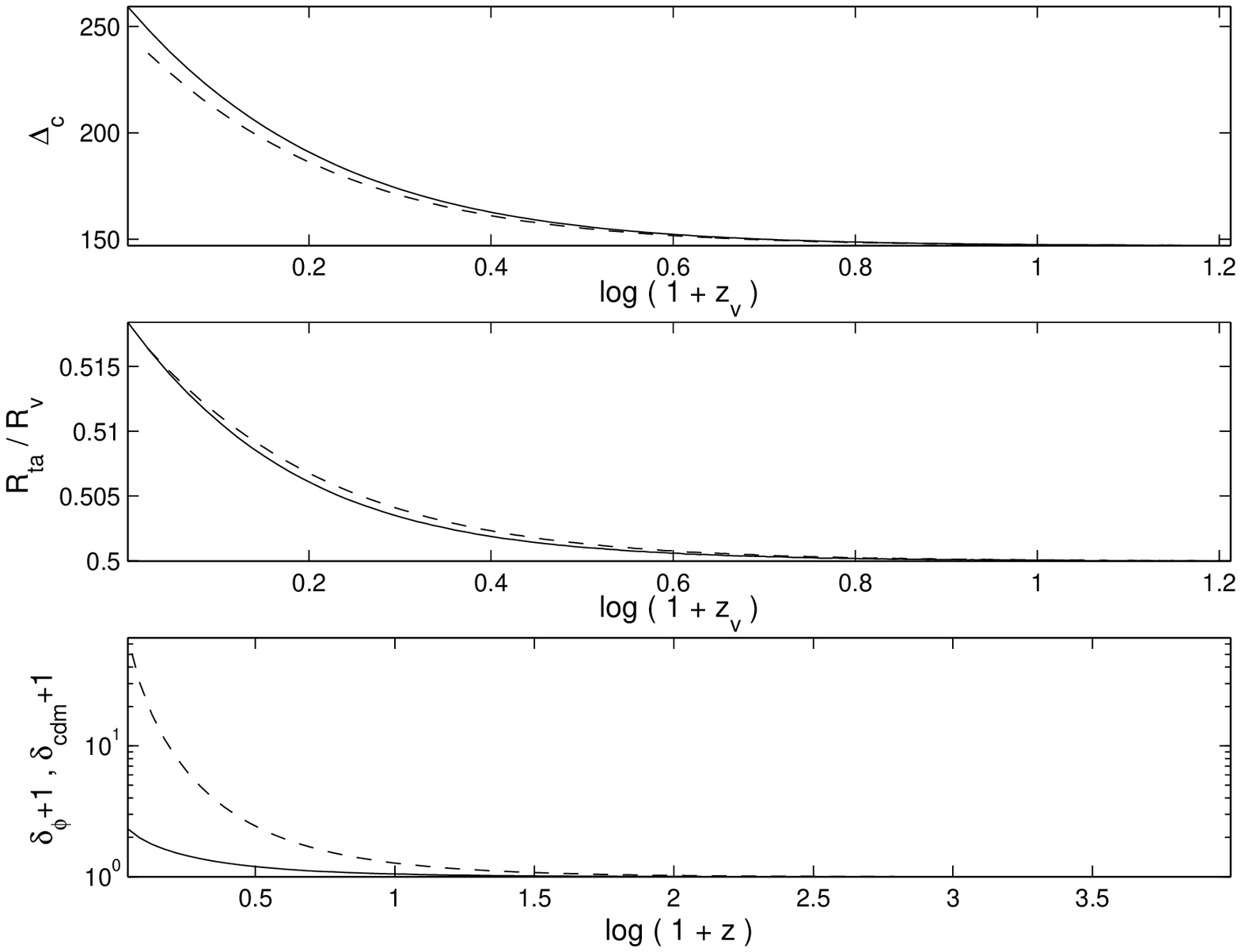}
\caption{{\protect\small {\textit{Model with constant equation of state $w=-0.8$.
The upper panel shows the predictions for $\Delta_c$ for the inhomogeneous case (dashed 
line) and homogeneous case (solid line). The middle panel shows the predictions 
for $R_{\rm V}/R_{\rm ta}$ for the inhomogeneous case (dashed line) and homogeneous
case (solid line). The lower panel shows the evolution of the density contrast in 
matter (dashed line) and dark energy (solid line) for the case of an inhomogeneous 
dark energy component (cluster virialises at $z=0$).}}}}
\label{wcdm}
\end{figure*}

In Figure \ref{wcdm} we have plotted the case for a dark energy model with constant 
equation of state $w=-0.8$. As it can clearly be seen, the predictions for 
$\Delta_{\rm c}$ and $R_{\rm v}/R_{\rm ta}$ for the cases of a homogeneous and 
an inhomogeneous dark energy component are very small in this case. This 
implies, that for models with constant equation of state and as long as the 
equation of state doesn't differ too much from $w=-1$, the fitting formulae 
presented in the literature are valid \cite{wang,weinberg}.

The reason for the differences observed is that in the highly non--linear regime 
the field will play an important role in determine the collapse time if $\Gamma=0$. 
Although at the time of virialisation the field is only subdominant, it will play an 
important role just before collapse. 
Thus, backreaction effects could significantly 
alter the predictions of the spherical collapse model. 

\subsection{Dependence on the Initial Conditions}
\begin{figure*}[htbp!]
\centering
\includegraphics[height=9cm]{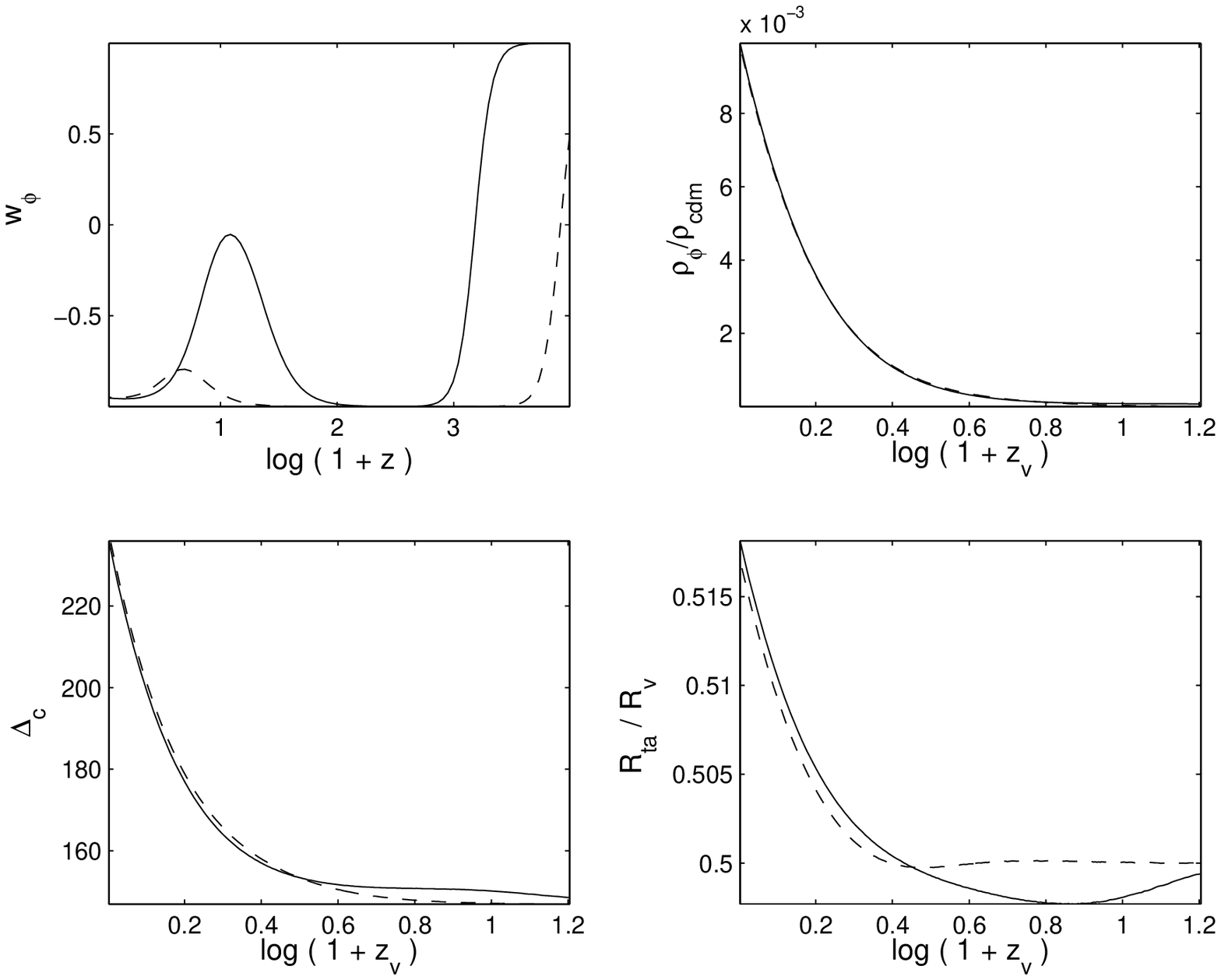}
\caption{{\protect\small {\textit{Dependence on the initial conditions for the 
double exponential potential \cite{copel} for the case of homogeneous dark energy. The solid and 
dashed lines show the results for different initial conditions.
Top left panel:
Evolution of $w_{\phi} $ in the background as a function of $\log (1+z)$. 
Top right panel: The radio $\rho_\phi/\rho_{\rm matter}$ inside the overdensity 
as a function of $\log (1+z_v)$.
Bottom left:  Evolution of $\Delta_{c}$ as a function of $\log(1+z_v)$. 
Bottom right: Evolution of $R_{ta}/R_v$  as a function of $\log(1+z_v)$.
 Whereas the 
evolution of the equation of state is different, the predictions for $\rho_\phi/\rho_{\rm cdm}$, 
$\Delta_c$ and $R_{\rm ta}/R_{\rm v}$ are only  weakly dependent on the initial 
conditions.}}}}
\label{diffininoclust}
\end{figure*}

We turn now our attention to the question, if the results of the 
spherical model depend on the initial conditions of the field $\phi$ 
(a discussion of initial conditions in the linear regime can be found in 
\cite{finelli,doran,bartolo}). 
We have run our code for different initial conditions and  
will discuss the case for an inhomogeneous ($\Gamma=0$) and homogeneous dark energy component 
independently. The case for a homogeneous dark energy component is shown in Figure 
\ref{diffininoclust}. In the left upper panel we show the evolution of the equation of 
state in the background for different initial values for $\phi$ at a
redshift of $10^4$. As it can be seen, the ratio 
$\rho_\phi /\rho_{\rm cdm,clus}$ is almost independent of the initial 
conditions chosen. The same is true for the predictions of 
$\Delta_{\rm c}$ and $R_{\rm ta}/R_{\rm v}$.

\begin{figure*}[htbp!]
\centering
\includegraphics[height=9cm]{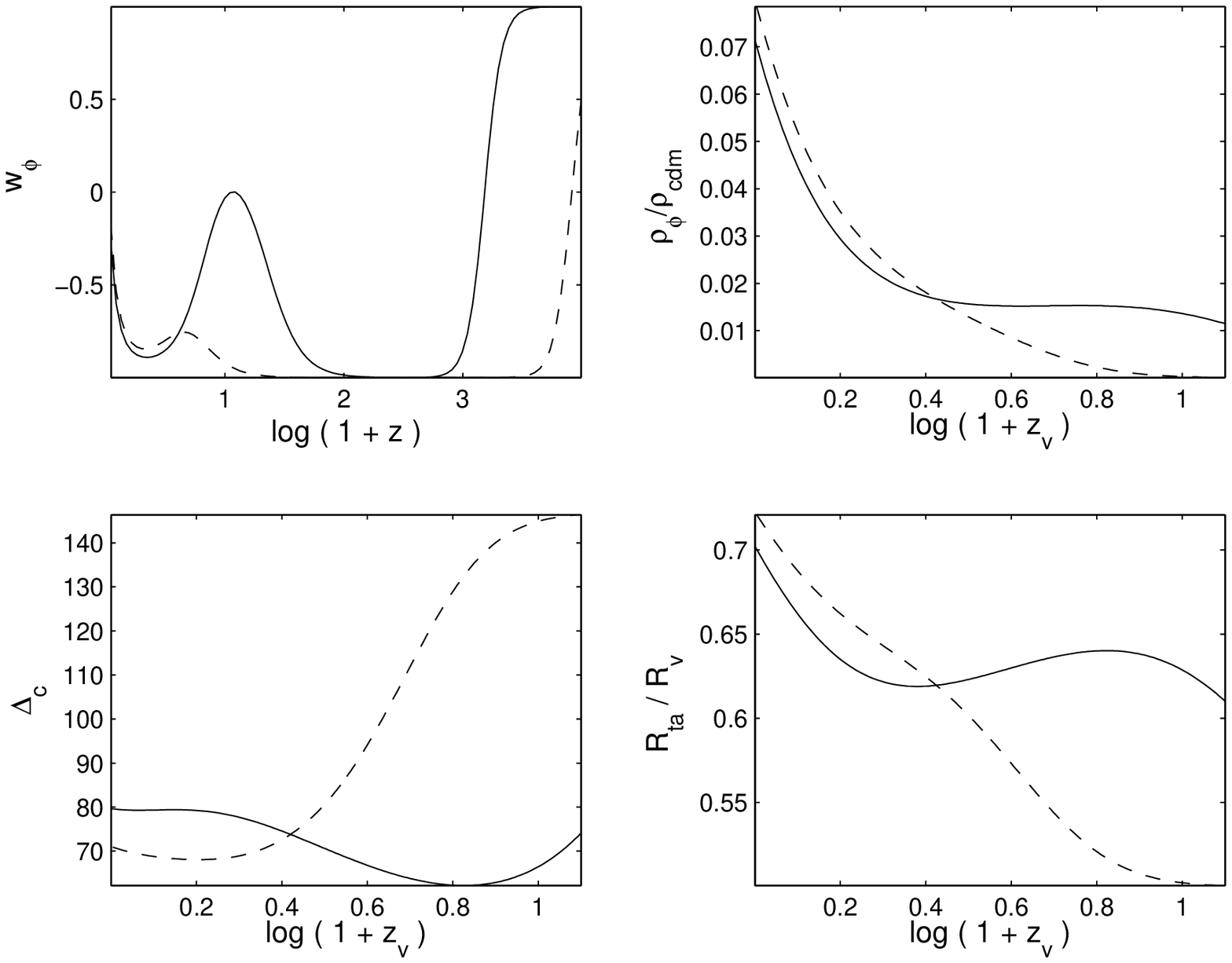}
\caption{{\protect\small {\textit{Dependence on the initial conditions for the 
double exponential potential \cite{copel} in the case for an inhomogeneous 
dark energy component ($\Gamma=0$).  The solid and 
dashed lines show the results for different initial conditions.
Top left panel:
Evolution of $w_{\phi} $ in the overdensity as a function of $\log (1+z)$. 
Top right panel: The radio $\rho_\phi/\rho_{\rm matter}$ inside the overdensity 
as a function of $\log (1+z_v)$.
Bottom left:  Evolution of $\Delta_{c}$ as a function of $\log(1+z_v)$. 
Bottom right: Evolution of $R_{ta}/R_v$  as a function of $\log(1+z_v)$.
It can clearly be seen that the predictions in this case 
depend on the initial conditions.}}}}
\label{diffiniclust}
\end{figure*}

The case for $\Gamma=0$ is shown in Figure \ref{diffiniclust}. We find that if quintessence 
becomes inhomogeneous together with dark matter in the strong non--linear regime, 
the predictions of the spherical collapse model depend generically 
on the initial conditions of the field. 

Different initial conditions lead to different time evolution of the
equation of state and the energy density of quintessence, both in the
background and inside the overdensity (depending on the value of
$\Gamma$). These two dark energy
properties will affect the dynamics of the overdensity through
 equations (\ref{rcluster}) and (\ref{generalpotential}), hence affecting the predictions for
 the non-linear density contrast and the virialisation radius of the
 overdensity.

\subsection{Comparison among the different models}

\begin{figure*}[htbp!]
\centering
\includegraphics[height=9cm]{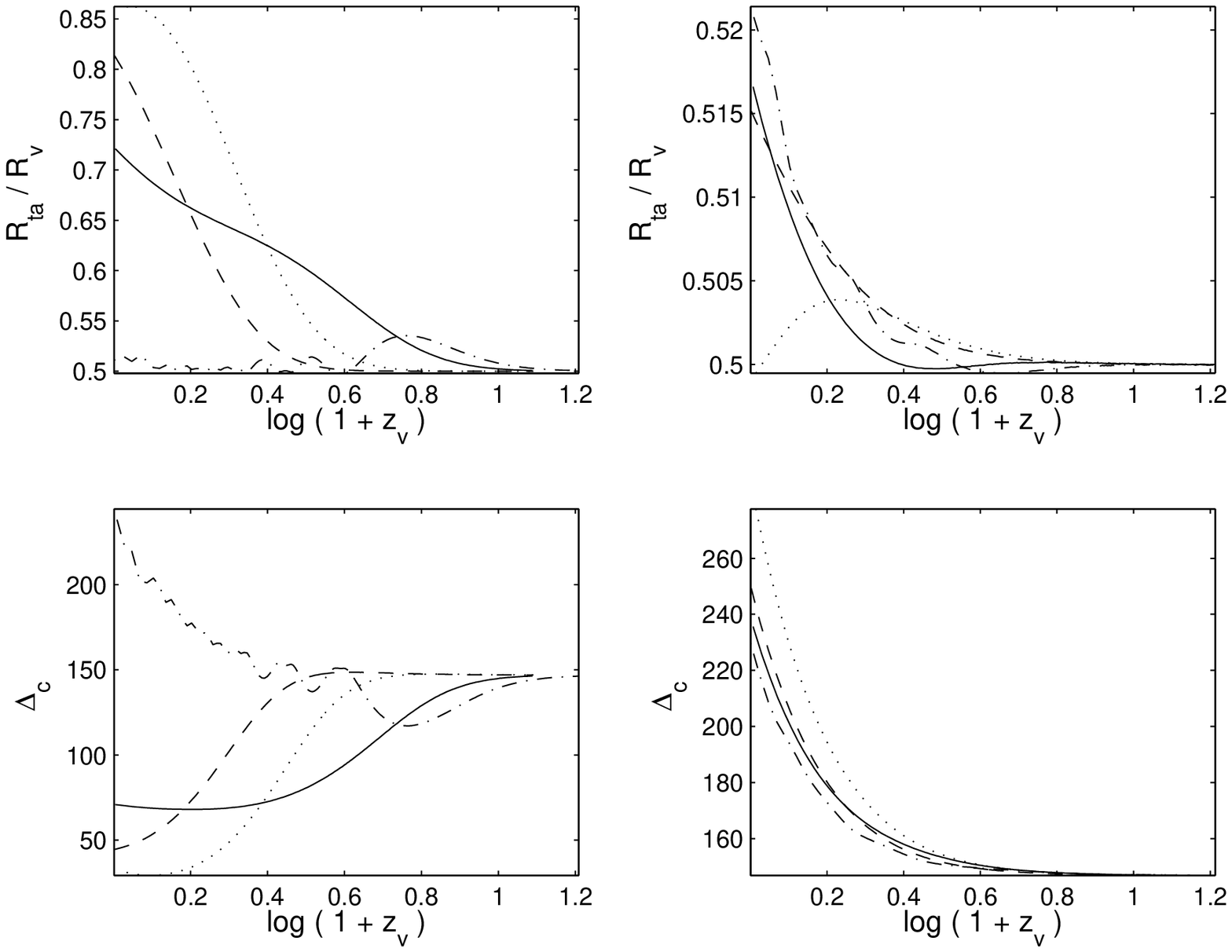}
\caption{{\protect\small {\textit{The predictions for $R_{\rm ta}/R_{\rm v}$ and 
$\Delta_c$ for the different quintessential potentials. The left panels show the 
case for an inhomogeneous dark energy component ($\Gamma=0$), whereas the 
right panel shows the results for the case of an homogeneous dark energy component.
It can be seen that the predictions depend on the potential used, although the 
dependence is weaker in the case of an homogeneous dark energy component. Solid line \cite{copel}, dashed-line \cite{steinhardt}, 
dashed-dotted line \cite{skordis}, dotted-line \cite{brax}}}}}
\label{comp}
\end{figure*}

What are the differences in the predictions for the four quintessence models? 
In Figure \ref{comp} we plot the predictions for $R_{\rm ta}/R_{\rm v}$ and  
$\Delta_{\rm c}$ for the different potentials with the different 
assumptions for $\Gamma$. As it can be seen from that Figure, the differences can 
be quite large if $\Gamma=0$. If the field is homogeneous throughout space, 
the differences are small. Again, the reason is that the dynamics of the overdensity 
(see equations (\ref{rcluster}) and (\ref{generalpotential})) strongly depends 
on the dark energy properties (Potential, $\Gamma$, etc).

\begin{figure}[Hhtbp!]
\centering
\includegraphics[height=7cm]{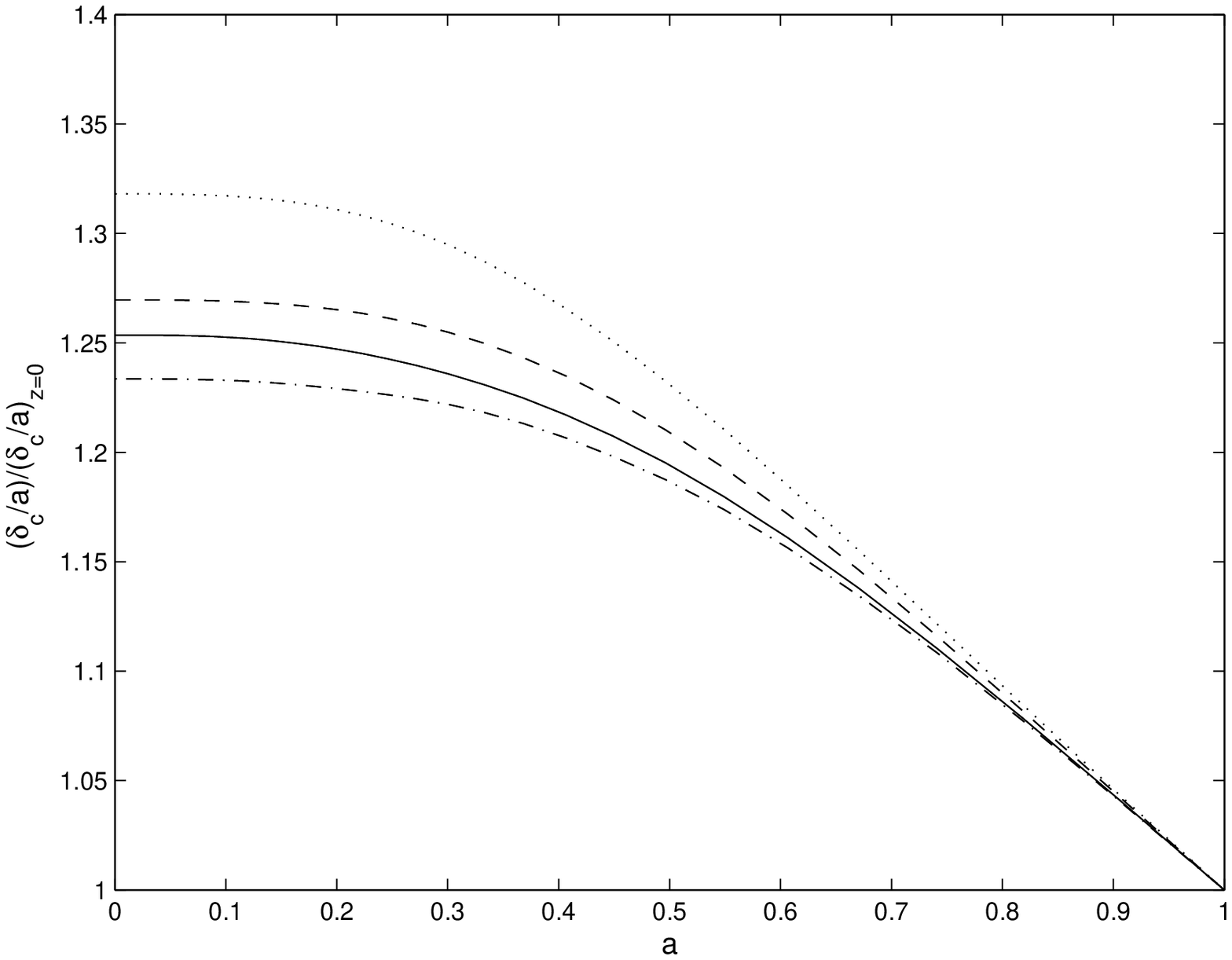}
\caption{{\protect\small {\textit{The linear growth factor for the different quintessential 
potentials. It can be seen that the growth factor depends on the potential used.
Solid line \cite{copel}, dashed-line \cite{steinhardt}, 
dashed-dotted line \cite{skordis}, dotted-line \cite{brax}}}}}
\label{growth}
\end{figure}

In Figure \ref{growth} we plot the predictions of the linear growth factor ($\delta_c /a$), 
normalised today, for the different potentials. As it can be seen, even in the linear 
regime there are differences between the four potentials. Such a behaviour was already 
noticed in \cite{maccio1}.

\subsection{Delayed collapse of dark energy}
\begin{figure}[Hhtbp!]
\centering
\includegraphics[height=8cm]{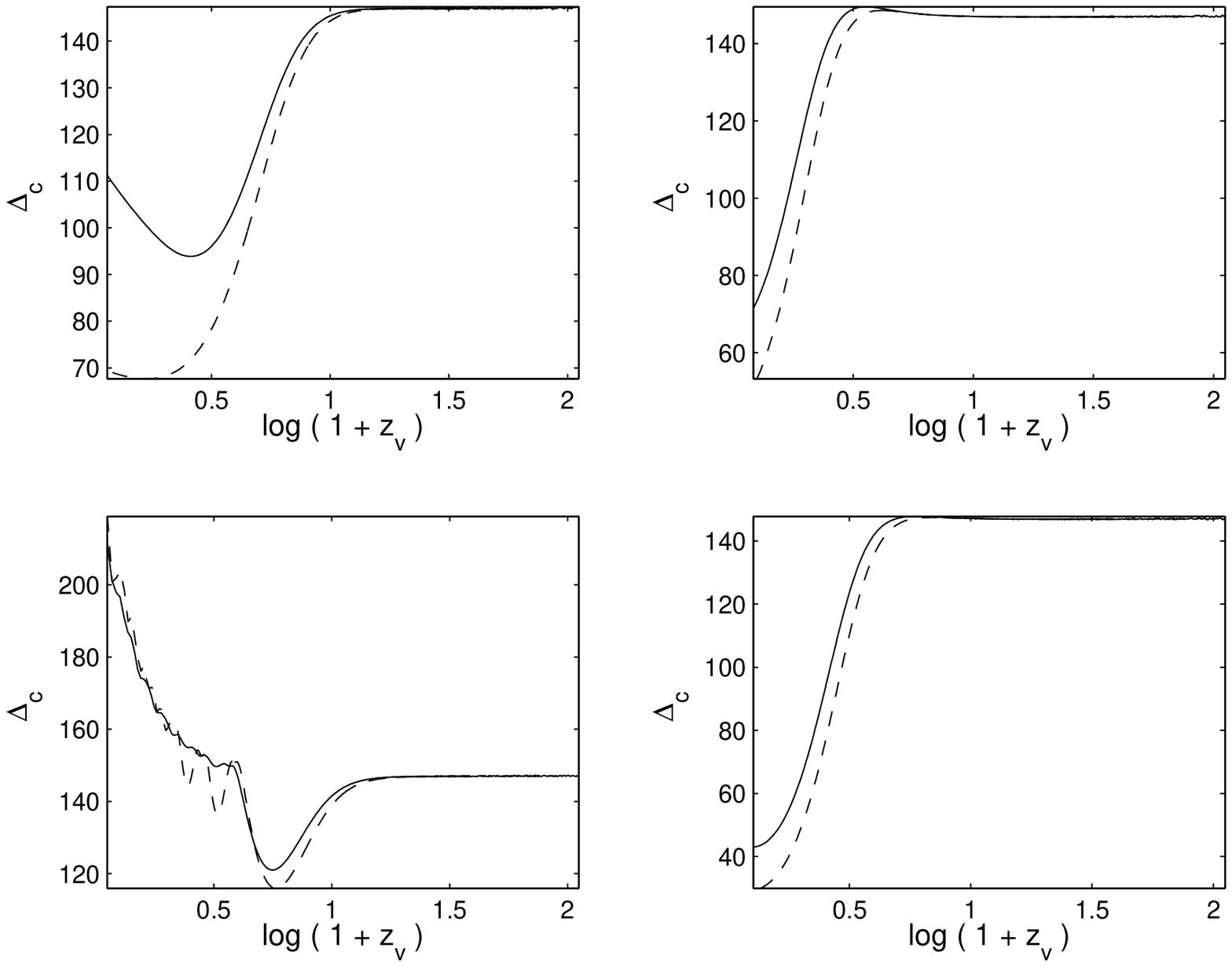}
\caption{{\protect\small {\textit{Predictions for $\Delta_c$ in the case of delayed collapse
of dark energy. In all four panels, the solid line corresponds to the delayed collapse situation, 
the dashed line to the case of $\Gamma=0$ both before and after 
turnaround. The upper left panel shows the predictions for 
the double exponential (\cite{copel}), the left upper panel for the model of \cite{steinhardt}, 
the lower left panel the model of \cite{skordis} and the lower right panel the model of 
\cite{brax}.}}}}
\label{turnaround}
\end{figure}

The results presented imply that if dark energy collapses together with dark matter, 
it has an important impact on the predictions of the spherical collapse model. Clearly, 
our assumption on the parameter $\Gamma$ so far is unrealistic. To gain more insight 
in how far our choices of $\Gamma$ changes the predictions of the spherical collapse 
model, we make now the assumption that the field inside the cluster is the same as 
in the background {\it until the turnaround}, from which on it collapses together with 
the dark matter. We will call this situation as the {\it delayed collapse of dark energy}. 
In the light of results obtained in the linear theory (see e.g \cite{ferreira}), 
this is a more realistic situation than studied up to now. 
The results for the precictions of $\Delta_c$ are shown in Fig. \ref{turnaround}. 

In Fig. \ref{turnaround} we show the results for $\Delta_c$ for all four potentials. 
The solid lines are the cases for delayed collapse of dark energy, 
whereas the dashed lines correspond to the case of full collapse. As it can be clearly seen, 
the predictions for $\Delta_c$ change. Note, however, that the predictions are still far off 
from the the results obtained in the case of a homogeneous scalar field. Of course, the field 
is in the linear regime, but it has still an important impact on the predictions of the spherical 
collapse model. 

\section{Conclusions}
In this paper we have studied the spherical collapse model in cosmologies 
with dark energy, provided by a scalar field (quintessence). Our aim 
was to study the model for different quintessence potentials and 
under different assumptions about the behaviour of the field during the highly
non--linear regime. Our results show, that the predictions of the 
spherical collapse model depend on the form of the potential, on the
initial conditions of the field and on the behaviour of the field 
in highly non--linear regions. If backreaction effects onto the field turn 
out to be important, these could significantly change the predictions of the 
spherical collapse model. 

We would like to point out that although the dark matter enters the highly 
non-linear regime first, perturbations in dark energy deviate only slightly from linear theory  
around virialisation. At the time of virialisation, the energy density of dark energy is always 
much smaller than the density of dark matter inside an overdensity. The main effect is that 
dark matter perturbations change the evolution of dark energy perturbations. 
The physical properties of clusters, such as the 
density contrast and virial radius strongly depend on clustering properties 
of dark energy (i.e. the parameter $\Gamma$), the quintessence 
potential and the initial conditions for the field fluctuations. The latter is 
only important, however, if the field collapses together with the dark matter.
On the other hand, we have found that if the dark energy equation of state is assumed to be 
constant, the differences between the homogeneous and inhomogeneous cases are 
small, at least if the equation of state $w$ does not differ too much from $w=-1$.
Thus, for constant equation of state the fitting formulae presented in the literature 
(see \cite{wang,weinberg}) do not change drastically even if inhomogeneities in the dark 
energy component are taken into account. 

Our results imply, that a better understanding of the behaviour of the 
quintessence field in highly non--linear regions is needed before the 
predictions of the spherical collapse model can be trusted. The field can 
influence the dynamics of the collapse at late times and thus changing 
predictions of the turnaround, virialisation and collapse times. In order 
to get better predictions, the boundary conditions between the outer and 
inner metric have to be understood better. Its likely that the spherical model just gives 
qualitative but not quantitative predictions. Nevertheless, an estimate 
of the function $\Gamma$ is needed, which can only be obtained from a 
general relativistic treatment. One way would be to develop a 
swiss cheese model in the case of a quintessential universe. This will be discussed 
in future work.

Our calculations have also implications for models in which the scalar field 
couples non--minimally to dark matter. In these models, the scalar field perturbations 
and the dark matter perturbations are coupled even in the linear regime and 
backreaction in the highly non--linear regime will certainly be important. 
A similar conclusion might hold for models based on tachyonic fields. 

To summarise, only when the energy flux of scalar field energy density 
$\Gamma$ out of a dark matter overdensity is known, the spherical 
collapse model is able to make firm predictions which can be used to 
make predictions for weak and strong lensing, the number density of 
clusters, etc or even for a help to use N--body simulations in dark energy 
models. In the case of a non--minimally coupled dark energy component, the 
predictions of the spherical collapse model will certainly be altered. 

\begin{acknowledgements}
We would like to thank P.G. Ferreira, J. Magueijo, K. Moodley, 
C. Skordis, J. Silk and D. Tocchini-Valentini, for helpful comments. 
DFM is supported by Funda\,c\~ao Ci\^encia e a Tecnologia. CvdB is 
supported by PPARC. 
\end{acknowledgements}

\end{document}